\documentstyle[11pt]{article}
\catcode`\@=11
\begin{document} \openup6pt

\title{ PRIMORDIAL BLACK HOLE PAIR CREATION PROBABILITY IN  MODIFIED
GRAVITATIONAL THEORY}

\author{  B. C. Paul\thanks{ e-mail :
bcpaul@iucaa.ernet.in }  and Dilip Paul \\
   Department of Physics, North Bengal University, \\
   Siliguri, Dist. Darjeeling, Pin : 734 013, India }

\date{}

\maketitle

\begin{abstract}

The  probability for quantum creation of an inflationary universe with a
 pair of black holes is computed in a modified gravitational theory. Considering
a gravitational action which includes  a cosmological constant ($\Lambda$) in
addition to $ \alpha\; R^{2} $ and  $ \delta \; R^{-1}$ terms, the probabilities  
have  been evaluated for two different kinds of spatial sections, one accommodating a pair of black holes and the other without black hole. We adopt a technique prescribed by Bousso and Hawking to calculate the above creation probability in a semiclassical approximation with Hartle-Hawking boundary
condition. Depending on the parameters in the action some new and physically interesting instanton solutions  are presented here which may play an important role in the  creation of the early  universe. We note that the probability of creation of a universe with a pair of black holes is strongly suppressed with a positive cosmological constant when   $\delta =  \frac{4 \Lambda^{2}}{3}$ for $\alpha > 0$ but it is more probable for $\alpha < - \frac{1}{6 \Lambda}$.  It is also found that instanton solutions are allowed without a cosmological constant in the theory provided $\delta < 0$.
\\
PACS No(s). : 04.70.Dy, 98.80.Bp, 98.80.Cq. \\

Key Words : Gravitational instanton, primordial black hole, quantum cosmology.

\end{abstract}

{\bf I. INTRODUCTION : }

It is now  understood from the information gathered from high redshift surveys of supernovae to Wilkinson Microwave Anisotropy Probe ( WMAP ) data that the present universe is accelerating [1]. The  universe might have also emerged from an inflationary era in the past predicted from COBE. In the last two decades a number of
literature  appeared  which explored early universe with an inflationary phase [2].
Inflation not only opens up new avenues in the interface of particle physics and
cosmology but also solves some of the outstanding problems not understood in the 
standard bigbang cosmology. Most successful  theory  developed so far is based on
the dynamics of scalar field with  a suitable potential which drives inflation. In order to explain present acceleration in the universe, various cosmological models  have been proposed with exotic fields that appeared with a new gravitational physics considering theories
other than scalar field, e.g., chaplygin gas [3], phantom fields [4] etc. It may be pointed out here that an alternative theory with a modification  to the Einstein-Hilbert action has been employed to obtain  early inflation instead of a scalar field. It is known that a modification to the  Einstein-Hilbert action with higher order terms
in curvature invariants  that are become effective in the  high curvature region 
 lead to modifications of the standard cosmology admitting a   de Sitter universe at early times
[5]. In the same way it is important to look for
modifications of the Einstein gravitational action with terms that 
might be  important at extremely low curvature  region  to explain the present
cosmic acceleration.  Recently, various cosmological models are proposed to explain the cosmic speed up with a
modification of the Einstein Hilbert action.  Carroll {\it et al.} [6]
considered  a gravitational action of the form
\begin{equation}
I = \frac{1}{2 \kappa^{2}} \int d^{4} x \sqrt{- g} \left[ R -  \; \frac{\mu^{4}}{R}
\right],
\end{equation}
where $R$ is the Ricci Scalar, $\kappa^{2} = 8 \pi G $, and $\mu$ is a mass scale of
the order of  Hubble scale, to construct such models of the universe. In the literature [7] a number of cosmological models are appeared with  the  above
action to construct  an alternative theory of dark energy and dark matter. However, in the theory de Sitter space is always unstable. It has been shown [8] that adding a $\alpha \; R^2$ term to the action allows a stable de Sitter space if the size of the baby universe is bigger than a size determined by $\mu^2$ in the action. Further, the theory containing both the $1/R$ and $R^2$ terms has better chance of passing solar system tests  than the theory based on simple $1/R$ corrections [9].
It is also important to explore the cosmological issues in the framework
of new theories which are important at the present epoch. The astrophysical objects,  for example black holes need  to be investigated in these
theoretical framework. The mass of these objects may be greater than
 the solar mass or even less. It is known in stellar physics that a 
blackhole is the ultimate corpse of a collapsing star when its mass exceeds 
twice the  mass of the Sun. Another kind of black holes are 
important in cosmology which might have formed due to quantum fluctuations 
of matter distribution in the early universe. These are termed as topological
 blackholes having mass many times smaller than the the solar mass.   In particular, black holes whose mass  ($ \sim 10^{15} g$) have life times comparable to the age of the universe so that they would be evaporating at present which is Hawking radiation [10]. It may provide sources of ultra high energy cosmic rays including extragalactic  gamma-rays [11] and other astrophysical phenomena whose mechanisms have not been solved completely. In addition to that, the relic black hole might contribute to the energy density of the present universe and solve the dark matter problem [12].
 Bousso and Hawking [13] 
(in short, BH) calculated the probability of the quantum creation of 
a universe 
with a pair of primordial black holes (in short, PBH) in  (3 + 1) dimensional universe. 
To compute the probablity measure, BH considered two different 
Euclidean space-time : (i) a universe with space-like sections with 
 $S^{3}$ - topology  and (ii) a universe with space-like section with 
$ S^{1} \times S^{2}$- topology, as is obtained in the Schwarzschild- de Sitter
 solution.
The first kind of spatial structure describes an 
inflationary (de Sitter) universe without  black hole while the second kind
 describes a Nariai 
universe [14], an inflationary universe with a pair of black holes.  
 Considering  a massive scalar field which provided an 
effective cosmological constant for a while through a slow-rolling potential 
(mass-term) BH studied the problem. 
 Chao [15] studied the creation of a primordial black hole and
 Green and Malik [16] studied the primordial black holes production during
 reheating. Paul {\it et al.} [17] following the approach of BH studied the
 probability
of creation of PBH including $R^{2}$-term in the Einstein action and found
that the probability is very much suppressed in the $R^{2}$ -theory. 
Paul and Saha [18] studied probability of creation of a pair of black hole with
higher order Lagrangian i.e., considering  higher loop contributions 
into the effective action that are higher than 
quadratic in $R$. Recently, the probability for quantum creation of
an inflationary universe with a pair of black hole in a  gravitational
theory  which includes  a cosmological constant
($\Lambda$) in addition to ($ \delta \; R^{- 1}$) term  is evaluated [19]. In this paper we investigate
the probability for primordial black hole pair creation in a modified gravity by
adding an extra term $\alpha R^{2}$ to the action given by (1) in the presence of a
cosmological constant. The above theory is important as it may be regarded as the synthesis of the theories one needs for early and late inflation, although it remains to be understood how the universe transits from early inflating phase to the present accelerating phase.

Consequently it is also important to study the effects of 
these terms to study quantum creation of a universe with a pair of PBH.  
We calculate the probabilities for the creation  of a universe with two
 types of topology namely, $ R \times S^{3} $ -  topology and 
$R \times S^{1} \times S^{2}$ - topology, where the later accommodates
a pair of primordial black holes. 
To calculate the probabilities for these spatial topologies,   we use a 
semiclassical approximation for the evaluation of  the 
Euclidean path integrals. The condition that a classical spacetime 
should emerge, to a good 
approximation, at a large Lorentzian time was selected by a choice of the
 path of the time parameter $\tau$ along the $\tau^{Re}$ axis 
from 0 to $\frac{\pi}{2H}$ and 
 then continues along the $\tau^{im}$ axis. 
The  wave-function of the universe in the semiclassical approximation  is 
given by
\begin{equation}
\Psi_{o} [ h_{ij}  , \Phi_{\partial  M} ]   \approx \sum_{n} A_{n}
 e^{- I_{n}}
\end{equation}
where the sum is over the saddle points of the path integral, 
and $I_{n}$
denotes the corresponding Euclidean action.  The probability measure of the
creation of PBH is
\[
P[ h_{ab} , \Phi_{\partial M} ] \sim  e^{ - 2 \;  I^{Re}}
\]
where $h_{ab}$ is the boundary metric and  $I^{Re}$  is the real part of
the action corresponding to the dominant saddle point, i.e. the classical
solution satisfying the Hartle-Hawking (henceforth, HH) boundary
conditions [20].  It was believed that all inflationary models lead a density parameter 
to $ \Omega_{o} \sim 1$ to a great accuracy. This view has been modified after it
was discovered  that there is a special class of inflaton effective 
potentials which
may lead to a nearly homogeneous open universe with  $ \Omega_{o} \leq 1$ 
at the present epoch. Cornish {\it et al.} [21, 22] studied the problem of 
pre-inflationary
homogeneity and outlined the possibility of creation of a small, compact,
 negatively
curved universe. We show that a universe with $S^{3}$-topology may give 
birth to
an open inflation similar to that obtained in {\it Ref.} 23.

The paper is organised as follows : in sec. II, the gravitational
 action for
a higher derivative gravity  is written and gravitational instanton solutions are presented
and
in sec. III, the action is used to estimate the relative probability for the 
two types of the universes one with PBH and the other without PBH. Finally, in sec. IV we give a brief discussion.

{\bf II. GRAVITATIONAL ACTION AND INSTANTON SOLUTIONS   
WITH OR 
WITHOUT A PAIR OF PRIMORDIAL BLACK HOLE : }

We consider Euclidean action which is given by
\begin{equation}
I_{E} =  - \frac{1}{16\pi} \int d^{4} x \sqrt{g} \;   f(R)  - 
\frac{1}{8\pi} \int_{\partial M} d^{3} x \sqrt{h}\;  K f'(R) 
\end{equation}
where $g$ represents  the 4-dimensional Euclidean metric, $\alpha$ and $\delta$ are dimensional
parameters, $\Lambda$ represents  the 4-dimensional cosmological constant and  $f(R)$ is a polynimial function in Ricci Scalar $R $.
The second term in the action is the gravitational surface term where $h_{ij}$ is the boundary metric and 
$K = h^{ij} K_{ij}$ is the trace of $K_{ij}$ , the second fundamental form of the
 boundary
$\partial M$ in the given metric $g $. The second term is the contribution from
 $\tau = 0$ back in the action. It vanishes for a universe with topology $S^{3}$, but gives a non vanishing contribution for topology
$S^{1} \times S^{2}$.

{\bf (i)   $S^{3}$-Topology, the de Sitter spacetime :}

In this section, we first derive the field equations from the  Eucliden action  
and then explore  instanton solutions in the modified gravitational theories.
 We now look for a solution 
with spacelike section having  $S^{3}$ -topology. The  corresponding 
four dimensional metric ansatz in this case  is given by
\begin{equation}
ds^{2} = d\tau^{2} + a^{2} (\tau) \left[ dx_{1}^{2} + sin^{2} x_{1} \;
d\Omega_{2}^{2} \right]
\end{equation}
where $a(\tau)$ is the scale factor of a four dimensional universe and
 $ d\Omega_{2}^{2} $ is a line element on the unit two sphere.
The scalar curvature is given by
\[
R = -  6 \left[ \frac{\ddot{a}}{a} + \left(
    \frac{\dot{a}^{2}}{a^{2}} - \frac{1}{a^{2}} \right) \right] .
\]
where an overdot denotes differentiation with respect to $\tau$. We rewrite 
the action (3), including the constraint through a Lagrangian multiplier 
$\beta$ and obtain 
\begin{equation}
I_{E} = -  \frac{\pi}{8}  \int 
         \left[ f(R) a^{3} - \beta \left( R + 6 \frac{\ddot{a}}{a} + 
6 \frac{\dot{a}^{2} -1}{a^{2}} \right) \right] d \tau 
- \frac{1}{8 \pi} \int_{\partial M} d^{3} x \sqrt{h} K f'(R).      
\end{equation}
Varying the action with respect to R, we determine
\begin{equation}
\beta = a^{3} f'(R).
\end{equation}
Substituting the above equation, treating $a$ and $R$ as independent 
variables we get 
\[
I_{E} = -  \frac{\pi}{8}  \int_{\tau = 0}^{\tau_{\frac{\pi}{2 H}} } 
         \left[ a^{3} f(R) - f'(R) \left( a^{3} R - 6  a \dot{a}^{2} - 6 a 
\right)  + 6 a^{2} \dot{a} \dot{R} f''(R)  \right] d \tau
\]
\begin{equation}
-  \frac{3 \pi}{4} \left[  \dot{a} a^{2} f'(R) \right]_{\tau = 0},
\end{equation}
here we have eliminated $\ddot{a}$ term in the action  by integration by parts. The
field 
equations are  now obtained by varying the action with respect to $a$  and
 $R$ respectively, giving
\begin{equation}
f''(R) \left[ R + 6  \frac{ \ddot{a}}{a}    + 
6 \frac{\dot{a}^{2} - 1}{a^{2}} \right]  = 0 ,
\end{equation}
\begin{equation}
2 f'''(R) \dot{R}^{2} + 2 f''(R) \left[ \ddot{R} + 2 \frac{\dot{a}}{a} 
\dot{R} \right] + f'(R) \left[ 4 \frac{ \ddot{a}}{a} + 2 
\frac{ \dot{a}^{2}}{a^{2}} - \frac{2}{a^{2}} + R \right] - f(R)  = 0.
\end{equation}
We now consider $f(R) = R - 2 \Lambda +  {\alpha}{R^2} + \frac{\delta}{R}  $ to solve Eqs. (8)-(9). An instanton solution is obtained which is given by
\begin{equation}
a = \frac{1}{H} \; \sin  H \tau 
\end{equation}
where $ R = 12 H^{2} $ and $H $  can be determined in terms of  the parameters $ \delta, \Lambda  $  from the constraint equation which emerged from eq.(9) :
\begin{equation}
48  H^{4} - 16 \Lambda H^{2} + \delta = 0. 
\end{equation}
We note following from the Eq. (11) : (i) A  real Hubble parameter is permitted for $\delta \leq \frac{4{\Lambda}^2}{3} $,  (ii) two classes of instantons are allowed for $0 <
\delta < \frac{4 \Lambda^{2}}{3}$, (iii) Instanton solution with $H = \sqrt{\frac{\Lambda}{6}} $ is obtained when $\delta =\frac{4{\Lambda}^2}{3} $,  (iv) Instanton solution with $H = \sqrt{\frac{\Lambda}{3}} $ is obtained when $\delta = 0$, and (v) Instanton solution is permitted even with 
 $\Lambda = 0$  when  $\delta < 0$.

It is evident that the instanton solution (10) obtained here satisfies the HH no boundary
conditions viz., 
$a(0) = 0 $, $ \dot{a} (0) = 1$. One can choose a path along the $\tau^{Re}$
axis to  $\tau = \frac{\pi}{2 H}$, the solution describes half of the 
Euclidean
de Sitter instanton $S^{3}$.  Analytic continuation of the metric (4) to 
Lorentzian region, $x_{1} \rightarrow \frac{\pi}{2} + i \sigma $, gives
\[
ds^{2} = d \tau^{2} + a^{2}(\tau) \left[ - d\sigma^{2} + \cosh^{2} \sigma
\;  d \Omega_{2}^{2} \right]
\]
which is a spatially inhomogeneous de Sitter like metric. However, if one 
sets $\tau = i t $ and  $ \sigma = i \frac{\pi}{2} + \chi $, the metric 
becomes
\begin{equation}
ds^{2} = -  dt^{2} + b^{2}(t) \;  [ d\chi^{2} + \sinh^{2} \chi
 \;
 d \Omega_{2}^{2} ]
\end{equation}
where $b(t) = - \; i \;  a( it )$. The line element (12) now describes  an open 
inflationary  universe.
The real part of the Euclidean action 
corresponding to the solution calculated by following the complex contour
of $\tau $ suggested by BH is given by
\begin{equation}
I^{Re}_{S^{3}} = - \frac{ \pi}{8} \left[ \frac{ 1728 \alpha H^{6} + 144 H^{4} - 24
\Lambda H^{2} +  \delta }{ 18 H^{6} } \right].
\end{equation}
With the chosen path for $\tau $,  the solution describes 
half the de Sitter instanton  with $S^{4}$ 
topology, joined to a real Lorentzian hyperboloid of
topology  $R^{1} \times S^{3}$. It can be joined to any boundary satisfying 
the condition 
$a_{\partial M} > 0$ . 
For $a_{\partial M} > H^{- 1}$ , the wave function oscillates and 
predicts a classical
space-time. 
We note the following cases :

(a) $S^{3}$ solution obtained by BH [13] is recovered when $\delta = 0$, $\alpha = 0$.  

(b)  $\delta = 0$, $\alpha \neq 0$,  corresponds to the
instanton solution obtained by Paul {\it et al.} [17].

(c) $\delta \neq  0$, $\alpha = 0$, $\Lambda \neq 0$, corresponds to the solution obtained by us recently [19] and the action reduces to
$I = - \frac{8 \pi}{\sqrt{- 3 \; \delta}}$ for a vanishing cosmological constant. It is evident here that a realistic solution is obtained with a negative $\delta$ only. 

(d)  $\delta \neq 0$, $\alpha \neq  0$ and $\Lambda = 0$, one obtains  action  which is
$I = - 4\;\pi \left[ 3\;\alpha + \frac{2}{\sqrt{- 3 \; \delta}}\right]$. But the above solution also demands $\delta < 0$.

(e)  $ \delta = \frac{4 \Lambda^{2}}{3}$,  we get the action which becomes $I = {-12\; \pi
\alpha} - \frac{2\;\pi}{\Lambda} $. It is interesting if $\alpha < 0$.

{\bf (ii) Spacelike sections $ S^{1} \times S^{2} $, the Nariai spacetime:}

We now look for a solution with spacelike sections $S^{1} \times
S^{2}$ as this topology accommodates a pair of black holes. The corresponding ansatz
for (1 + 1 + 2) dimensions is given by
\begin{equation}
ds^{2} = d \tau^{2} + a^{2}(\tau) \; dx^{2} + b^{2}(\tau) \;  d \Omega_{2}^{2} 
\end{equation}
where $a( \tau ) $ is the scale factor of $S^{1}$-surface and $b( \tau )$ is the
scale factor of the two sphere. The metric for the two-sphere is  given by the metric 
\[
d\Omega_{2}^{2} = d \theta^{2} + sin^{2} \theta \; d \phi^{2} 
\]
The scalar curvature is given by
\begin{equation}
R = - \left[ 2 \frac{\ddot{a}}{a} +  4 \frac{\ddot{b }}{b}
     + 2 \left( \frac{\dot{b}^{2}}{b^{2}} - 
   \frac{1}{b^{2}} \right) + 4 \frac{\dot{a} \dot{b}}{a b} \right] .
\end{equation}
The Euclidean action (3) becomes
\[
I_{E} = - \frac{\pi}{2} \int
 \left[  f(R) a b^{2} - \beta \left(R + 2 \frac{\ddot{a}}{a} + 4
\frac{\ddot{b}}{b} + 4 \frac{\dot{a} \dot{b}}{ab} + 2 
\frac{\dot{b}^{2}}{b^{2}} - \frac{2}{b^{2}}\right) \right] d\tau
\]
\begin{equation}
 - \frac{1}{8 \pi} \int_{\partial M} \sqrt{h} \;  d^{3}x \;  K f'(R).
\end{equation}
One can determine $\beta$ as is done before and obtains 
\[
I_{S^{1} \times S^{2}} = - \frac{\pi}{2} 
\int_{\tau = 0}^{\tau_{\partial M}} 
 \left[  f(R) - f'(R) \left( R - 4  \frac{\dot{a} \dot{b}}{ab} - 
2 \frac{\dot{b}^{2}}{b^{2}} - \frac{2}{b^{2}}\right) + 2 f''(R)  
\dot{R} \left( \frac{\dot{a}}{a} + 2 \frac{\dot{b}}{b} \right) \right] 
\]
\begin{equation}
a b^{2} d \tau - \pi \left[ \left( \dot{a} b^{2}  + 2 a b \dot{b} \right) f'(R) 
\right]_{\tau = 0}.
\end{equation}
Variation of the action with respect to $R $, $a$ and $b$
 respectively are given by
the following  equations
\begin{equation}
f''(R) \left[  R + 2 \frac{\ddot{a}}{a}
  + 4 \frac{\ddot{b}}{b} + 4 \frac{\dot{a} \dot{b}}{ab}  
+ 2 \frac{\dot{b}^{2}}{b^{2}} - \frac{2}{b^{2}} \right] = 0,
\end{equation}
\begin{equation}
2  f'''(R) \dot{R}^{2} + 2 f''(R) \left[ \ddot{R} + 2 \dot{R} 
\frac{\dot{b}}{b}
\right] + f'(R) \left[ R + 4 
\frac{\ddot{b}}{b}  +  
2 \frac{ \dot{b}^{2} - 1}{b^{2}} \right] - f(R)  = 0,
\end{equation}
\[
2 f'''(R) \dot{R}^{2} + 2 f''(R) \left[ \dot{R} \left(\frac{\dot{a}}{a} +
\frac{\dot{b}}{b}\right)  + \ddot{R} \right] + f'(R)  
\left[  R + 2  \frac{\ddot{a}}{a} +
 2 \frac{\ddot{b}}{b}  + 2 \frac{\dot{a} \dot{b}}{ab} \right] 
\]
\begin{equation}
- \; f(R)   = 0.
\end{equation}
Let us now consider $f(R) = R - 2 \Lambda + {\alpha \; R^{2}} + \frac{\delta}{R}  $.
Eqs. (18)-(20) admit an instanton solution  which is given by
\[
a = \frac{1}{H_{o}} \; sin ( H_{o} \tau ) , \; \; \;
b =  H_{o}^{- 1},
\]
\begin{equation}
R = 4 H_{o}^{2}, 
\end{equation}
where $H_{o}$ satisfies the constraint equation 
\begin{equation}
16  H_{o}^{4} - 16 \Lambda  H_{o}^{2} + 3 \delta  = 0.
\end{equation}
 We note the following : (i) when $ 0  < \delta < \frac{4 \Lambda^{2}}{3}$, two classes of instanton solutions are permitted, (ii) $\delta = 0$, instanton solution with   $H_{o}^{2} = \Lambda$ is obtained, (iii) $\delta =  \frac{4}{3} \Lambda^{2}$ leads to an instanton solution with $H_{o}^{2} = \frac{\Lambda}{2}$, (iv)  $\Lambda  = 0$, permits instanton with  $H_{o}^{2} = \sqrt{ - \; \frac{3 \delta}{16} }$ when  $\delta < 0$.

The above instanton solution  (21) satisfies the HH boundary conditions
$a(0) = 0 $, $ \dot{a} (0) = 1,  b(0) = b_{o} $, $ \dot{b} (0) = 0$.
Analytic continuation of the metric (14) to Lorentzian region, i.e.,
$\tau \rightarrow it $ and $ x \rightarrow \frac{\pi}{2} + i \sigma $ yields
\begin{equation}
ds^{2} = -  dt^{2} + c^{2}(t) \; d\sigma^{2} + H_{o}^{-2} \; d \Omega_{2}^{2},
\end{equation}
where $c(t) = - \;  i \;  a( i t )$. In this case the analytic continuation of time
and space
do not give an open inflationary universe. The corresponding Lorentzian 
solution
is given by
\[
a(\tau^{Im}) |_{\tau^{Re} \;  = \frac{\pi}{2H_{o}}} = H_{o}^{-1} \cosh H_{o}
\tau^{Im} ,
\]
\begin{equation}
b(\tau^{Im}) |_{\tau^{Re} \;  = \frac{\pi}{2H_{o}}} = H_{o}^{-1} 
\end{equation}
Its space like sections can be visualised as  three spheres of radius $H_{o}^{-1}$
with a {\it hole} of radius $b = H_{o}^{-1}$ punched through the north and south
poles. 
The
physical interpretation of the solution is that of two - spheres containing
two black holes at opposite ends. The black holes have the radius $H_{o}^{-1}$ which
accelerates away from each other with the expansion of the universe. This describes
half of a Lorentzian Nariai universe. The real part of the action can be
determined now following the contour suggested  by BH [13], and it is given by
\begin{equation}
 I^{Re}_{S^{1} \times S^{2}} = - \frac{\pi}{8  H_{o}^{2}} \left[ {64 \alpha
  H_{o}^{6}} + {16  H_{o}^{4}} -{8 \Lambda H_{0}^{2}} + { \delta} \right].
\end{equation}
The solution given by eq.(24) describes a universe with two black holes at the poles
of a two sphere. It may be pointed out  here that the contribution of the integrand
in the action (17) for the instanton vanishes and the non-zero contribution of the action here
arises just from the boundary term only. We note the following :

(a)  $\delta = 0$, $\alpha = 0$, corresponds to the action obtained by BH [13].

(b)  $\delta = 0$,  $\alpha \neq 0$ corresponds to  $I
= {-8 \pi \alpha} - {\frac{\pi}{\Lambda}}$, it reduces to a  case discussed by Paul {\it et al.} [17].

(c)   $\delta \neq 0$, $\alpha = 0$, $\Lambda = 0$, the action is 
$I = - \frac{16 \pi}{3 \sqrt{- 3 \; \delta}}$, it is real for $\delta < 0 $ , which
corresponds to the result obtained by us [19].

(d) $\delta \neq 0$, $\alpha \neq 0$ and $\Lambda = 0$ corresponds to an  action 
which is 
\[                                                                                  
            I^{Re}_{S^{1} \times S^{2}} = - \; { \frac{8 \pi}{3}} \left( {3 \alpha} +
\frac{2}{ \sqrt{- 3 \; \delta}}\right) = {\frac{2}{3}} I^{Re}_{S^{3}}.
\]
In this case a physically realistic solution is obtained for  $\delta < 0$.

(e)  $\delta = \frac{4 \Lambda^{2}}{3}$, the action becomes  
\[
I^{Re}_{S^{1} \times S^{2}} = - \; \frac {4 \pi}{3} \left(6 \alpha + \frac{1}{\Lambda} \right) =  \frac{2}{3}  I^{Re}_{S^{3}}.
\] 

{\bf III. THE  PROBABILITY FOR THE PRIMORDIAL BLACK HOLES PAIR CREATION  :}

In the previous section we have calculated the actions for inflationary 
universe with or without a pair of black holes. We now compare the
 probability
measure  in the two cases.
The probability for creation of a  de Sitter universe 
is determined from the action (7).  The probability for
nucleation of an inflationary universe without  a pair of Black holes is given by
\begin{equation}
P_{S^{3}} \sim \; e^{ \frac{\pi}{9} \left[{ \frac {216 \alpha  H^{4} + 12 H^{2}-
\Lambda }{ H^{4}}}\right] }.
\end{equation}
However for an inflationary universe with a pair of black holes the 
corresponding probability of nucleation can be obtained from the
action (17).  The corresponding probability is
\begin{equation}
P_{S^{1} \times S^{2}} \sim e^{ \frac{ 2 \pi}{3} \left[{\frac
 {24 \alpha  H_{o}^{4} + 4 H_{0}^{2}- \Lambda}{ H_{0}^{4}}} \right] }.
\end{equation}
For simplicity to compare the probabilities we consider spacial cases  :

(a)   The probabilities obtained by 
Bousso and Hawking [13] is recovered when both $\delta $ and  $\alpha$ are set zero. The probabilities are
\begin{equation}
P_{S^{3}} \sim e^{\frac{3 \pi}{\Lambda}}, \; \; \; P_{S^{1} \times S^{2}} \sim e^{\frac{2 \pi}{\Lambda}}.  
\end{equation}
In this case the probability for a universe with PBH is less than that without PBH. The solution cannot admit a negative cosmological constant.

(b)   The probabilities evaluated by Paul {\it et al.} [17] are recovered  here with  $\delta =  0$. The probabilities are
\begin{equation}
P_{S^{3}} \sim e^{{\frac{3 \pi}{\Lambda}} + {24 \pi \alpha}}, \; \; \; P_{S^{1} 
\times S^{2}} \sim e^{{\frac{2 \pi}{\Lambda}} + {16 \pi \alpha}}.
\end{equation}
In this case with a positive cosmological constant and $\alpha > 0$, the de Sitter
universe is more probable. However for $\alpha < -\frac {1}{8 \Lambda}$, the
probability for a universe with PBH is more. However, the case $\alpha < 0$ leads to
a classical instability in $R^{2}$ theory.

(c) For $\delta \neq  0$ and $\alpha = 0$, the solutions obtained in {\it Ref.} 19 are recovered.  There are two branches of solutions admitting two different instanton solutions. In one case  
the probabilities  are given by
\begin{equation}
P_{S^{3}} \sim e^{\frac{6 \pi}{ \left( \Lambda +  \sqrt{ \Lambda^{2} - \frac{3 \delta}{4}} \right)^{3}} \left[  2  \Lambda  \left(  \Lambda  + \sqrt{ \Lambda^{2} - \frac{3 \delta}{4}} \right)  - \delta \right]}
,  \; \; \; 
P_{S^{1} \times S^{2}} 
\sim e^{\frac{4 \pi}{ \left( \Lambda +  \sqrt{ \Lambda^{2} - \frac{3 \delta}{4}} \right)^{3}} \left[  2  \Lambda  \left(  \Lambda  + \sqrt{ \Lambda^{2} - \frac{3 \delta}{4}} \right)  - \delta \right]}
\end{equation}
for $\delta < \frac{4 \Lambda^{2}}{3}$, and in the other case the probabilities are 
\begin{equation}
P_{S^{3}} \sim e^{\frac{6 \pi}{ \left( \Lambda -  \sqrt{ \Lambda^{2} - \frac{3 \delta}{4}} \right)^{3}} \left[  2  \Lambda  \left(  \Lambda  - \sqrt{ \Lambda^{2} - \frac{3 \delta}{4}} \right)  - \delta \right]}
,  \; \; \; 
P_{S^{1} \times S^{2}} 
\sim e^{\frac{4 \pi}{ \left( \Lambda -  \sqrt{ \Lambda^{2} - \frac{3 \delta}{4}} \right)^{3}} \left[  2  \Lambda  \left(  \Lambda  - \sqrt{ \Lambda^{2} - \frac{3 \delta}{4}} \right)  - \delta \right]}
\end{equation}
for $0 < \delta < \frac{4 \Lambda^{2}}{3}$.
It is evident from eqs. (30) and (31) that the creation of  a universe without PBH is more probable than with PBH
for  $\delta < \frac{4 \Lambda^{2}}{3} \; $ with a positive cosmological constant. It admits negative $\delta$ also. However,  the probability for creation of a universe with a pair of PBH is favoured in the  first case when one considers a negative cosmological constant. In the second case, the probability of creation of a universe with a pair of PBH is favoured with a positive cosmological constant for $0 < \delta < \Lambda^{2}$, but it is suppressed  for 
$ \Lambda^{2} < \delta < \frac{4 \Lambda^{2}}{3}$. In the second case no instanton solution exists with a negative cosmological constant.

(d) For $\delta = \frac{4 \Lambda^{2}}{3}$ and $\alpha \neq 0 $, the probabilities are given by 
\begin{equation}
P_{S^{3}} \; \sim  \; e^{{\frac{4 \pi}{\Lambda}} + {24 \pi \alpha}},  \; \; \; \; 
P_{S^{1} \times  S^{2}} \; \sim  \;  e^{{\frac{8 \pi}{ 3 \Lambda}} + {16 \pi
\alpha}}.
\end{equation}
In this case a positive cosmological constant and $\alpha > 0$ leads to a de Sitter
universe creation more probable than a universe with a pair of black holes. However, for $\alpha < -\frac{1}{6 \Lambda}$, we note that
creation of a universe with a PBH is  more probable. 

(e)  For a theory without cosmological constant,  the evaluated probabilities are 
\begin{equation}
P_{S^{3}} \; \sim  \; e^{24 \pi \alpha + \frac{16 \pi}{\sqrt{- 3 \delta}}},  \; \; 
\; \; P_{S^{1} \times S^{2}} \; \sim  \; e^{16 \pi \alpha + \frac{32 \pi}{ 3 \sqrt{-
3 \delta}}}.
\end{equation}
It is evident that  negative values for the parameter $\delta$  are permitted to obtain a physically relevant solution. For $\delta < 0$, we note that a universe without a PBH is more probable for positive values of the prameter $\alpha$, however, it is less probable if one considers a negative $\alpha$ satisfying the inequality $\alpha < - \frac{2}{3} \frac{1}{\sqrt{- 3 \delta}}$, determined by a negative $\delta$.

{\bf IV. DISCUSSIONS : }

In this letter we  estimated and compared the probability for creation of a universe with a primordial
black hole (PBH) pair  with a universe without such PBH pair considering  a modified theory of gravity. In
section II, we obtained  gravitational instanton solutions in  the two 
cases : (i)  a universe with $R \times S^{3}$ - topology  and  (ii) a universe with 
$ R \times S^{1} \times S^{2} $ - topology respectively. The 
Euclidean  action is then evaluated corresponding to the instanton solutions using BH technique [13]. We
found  that the probability of a
 universe
with $R \times S^{3} $ topology turns out to be lower than a universe
 with topology $ R \times S^{1} \times S^{2} $ in some specified restricted
conditions. It may be mentioned here that one gets a regular instanton in 
$S^{4} $ - topology
if there is no black hole. The existence of black hole  
restricts such a regular topology. 
The results obtained here on the probability of creation of a universe with a  pair
of primordial black holes are found to be strongly 
suppressed depending on the parameter $\delta$ decided by the cosmological constant
in some cases which are determined here. It is noted that the probability for a universe with PBH is strongly suppressed for small values of $\delta$ lying in the sector $0 < \delta <  \frac{4 \Lambda^{2}}{3}$ in the presence of a positive cosmological constant.  We note an interesting solution here in the
framework of the modified gravitational action with inverse power of $R$ -theory,
which admits  de Sitter instantons with $S^{3}$ and $S^{1} \times S^{2}$ topologies
even without a cosmological constant. However, a negative $\delta$ is required if one considers a theory without a cosmological constant. In this case the action is similar to that
considered by Carroll {\it et al.} [6] to obtain an accelerating universe today. We also note another case where a de Sitter universe is less probable if $\alpha < - \frac{1}{6 \Lambda}$. The result obtained earlier  with curvature square term in the action in Ref. [17] given by eq. (29) now can be compared with eq. (32). It is evident that although the effect of inverse square term in the action enhances the probability, the condition to get a universe with a PBH remained unaltered in the theories here.
It is interesting to note here that analytic continuation of a 
$R \times S^{3}$ metric considered here
to Lorentzian region leads to an open 3 - space.  One
obtains Hawking-Turok [23] type open inflationary universe in this case.
In the 
other type of topology  an open inflation section of the universe  is not permitted.
A detail study of an open inflationary universe will be discussed elsewhere.
Thus in a modified Lagrangian with inverse power in  $R$-theory, quantum creation of
PBH seems 
to be suppressed in the minisuperspace cosmology for some values of the 
parameters in the action, which are determined here.
Another new result obtained here is that gravitational instanton solution may be
obtained even with a negative cosmological constant which is not permitted in the
case considered by BH [13].

\vspace{0.5 in}

{\bf {\it Acknowledgement  :}}
BCP would like to thank Inter-University Centre for Astronomy and Astrophysics (IUCAA), Pune for awarding  Visiting Associateship of IUCAA. Authors like
to thank S. Mukherjee for providing facilities of the IUCAA Reference Centre and Physics Department at
North Bengal University to carry out a part of the work.  This  work is partially
supported by University Grants Commission, New Delhi.

\end{document}